\documentclass[reprint, amsmath,amssymb,superscriptaddress, aps, prl,]{revtex4-2}
\usepackage{xcolor}
\usepackage{placeins}
\usepackage{soul}
\usepackage{graphicx}
\usepackage{dcolumn}
\usepackage{bm}
\usepackage{gensymb}
\usepackage{textcomp}
\usepackage{makecell}
\usepackage{multirow}
\usepackage{nicematrix}
\begin{document}
\title{Graphene-perovskite fibre photodetectors}
\author{S. Akhavan}
\thanks{present address: Institute for Materials Discovery, University College London, Torrington Place, London, WC1E 7JE, UK}
\affiliation{Cambridge Graphene Centre, University of Cambridge, JJ Thompson Avenue, Cambridge CB3 0FA, UK}
\author{A. Taheri Najafabadi}
\thanks{present address: Faculty of Engineering and Science, University of Greenwich, Central Avenue, Chatham Maritime, Kent, ME4 4TB, UK}
\affiliation{Cambridge Graphene Centre, University of Cambridge, JJ Thompson Avenue, Cambridge CB3 0FA, UK}
\author{S. Mignuzzi}
\affiliation{Cambridge Graphene Centre, University of Cambridge, JJ Thompson Avenue, Cambridge CB3 0FA, UK}
\author{M. Abdi Jalebi}
\thanks{present address: Institute for Materials Discovery, University College London, Torrington Place, London, WC1E 7JE, UK}
\affiliation{Cavendish Laboratory, University of Cambridge, JJ Thompson Avenue, Cambridge CB3 0HE, UK}
\author{A. Ruocco}
\thanks{Optical Networks Group, University College London, London, WC1E 6BT, UK}
\affiliation{Cambridge Graphene Centre, University of Cambridge, JJ Thompson Avenue, Cambridge CB3 0FA, UK}
\author{I. Paradisanos}
\affiliation{Cambridge Graphene Centre, University of Cambridge, JJ Thompson Avenue, Cambridge CB3 0FA, UK}
\author{O. Balci}
\affiliation{Cambridge Graphene Centre, University of Cambridge, JJ Thompson Avenue, Cambridge CB3 0FA, UK}
\author{Z. Andaji-Garmaroudi}
\affiliation{Cavendish Laboratory, University of Cambridge, JJ Thompson Avenue, Cambridge CB3 0HE, UK}
\author{I. Goykhman}
\affiliation{Cambridge Graphene Centre, University of Cambridge, JJ Thompson Avenue, Cambridge CB3 0FA, UK}
\affiliation{Technion - Israel Institute of Technology, Haifa 3200003, Israel}
\author{L. G. Occhipinti}
\affiliation{Cambridge Graphene Centre, University of Cambridge, JJ Thompson Avenue, Cambridge CB3 0FA, UK}
\author{E. Lidorikis}
\affiliation{Department of Materials Science and Engineering, University of Ioannina, Ioannina 45110, Greece}
\author{S. D. Stranks}
\affiliation{Department of Chemical Engineering and Biotechnology, University of Cambridge, Philippa Fawcett Drive, Cambridge CB3 0AS, UK}
\author{A. C. Ferrari}
\affiliation{Cambridge Graphene Centre, University of Cambridge, JJ Thompson Avenue, Cambridge CB3 0FA, UK}
\begin{abstract}
The integration of optoelectronic devices, such as transistors and photodetectors (PDs), into wearables and textiles is of great interest for applications such as healthcare and physiological monitoring. These require flexible/wearable systems adaptable to body motions, thus materials conformable to non-planar surfaces, and able to maintain performance under mechanical distortions. Here, we prepare fibre PDs combining rolled graphene layers and photoactive perovskites. Conductive fibres ($\sim$500$\Omega$/cm) are made by rolling single layer graphene (SLG) around silica fibres, followed by deposition of a dielectric layer (Al$_{2}$O$_{3}$ and parylene C), another rolled SLG as channel, and perovskite as photoactive component. The resulting gate-tunable PDs have response time$\sim$5ms, with an external responsivity$\sim$22kA/W at 488nm for 1V bias. The external responsivity is two orders of magnitude higher and the response time one order of magnitude faster than state-of-the-art wearable fibre based PDs. Under bending at 4mm radius, up to$\sim$80\% photocurrent is maintained. Washability tests show$\sim$72\% of initial photocurrent after 30 cycles, promising for wearable applications.
\end{abstract}
\maketitle
\section{\label{sec:level1}Introduction}
Electronics on fibres and textiles is of interest for a range of applications, including healthcare (e.g. monitoring heart rate)\cite{khan2016monitoring}, wearable displays\cite{choi2015wearable}, energy harvesting\cite{wang2015flexible, nathan2012flexible} and storage\cite{hu2010stretchable}. The current vision for electronic-textiles is for them to be an integral part of everyday outfits\cite{zeng2014fiber, grancaric2018conductive}, and remain functional after washing\cite{lou2020highly}. One of the most common approaches to test the washability of e-textiles was developed by the American Association of Textile Chemists and Colorists (AATCC)\cite{AATCC}. One cycle in AATCC is equivalent to 5 regular-machine (household) washing cycles\cite{rotzler2021washability}, achieved by the addition of steel or rubber balls to the washing container\cite{rotzler2021washability}, with agitator speed 179$\pm$2spm and spin speed 645$\pm$15rpm\cite{AATCC}. Washability tests are performed for at least 20 cycles\cite{rotzler2021washability}, with each cycle being 45min at 40$\degree$C, with$\sim$0.37\% volume of detergent (AATCC standard reference detergent 124\cite{AATCC}), and 10 stainless steel balls (6mm) or load ballast (130$\pm$10g for cotton) to mimic abrasion\cite{AATCC, karim2017scalable}. The analysis of washability depends on the intended application, the expected frequency of cleaning, and the total number of cleaning cycles during the tested product's lifetime. E.g., sports clothing will be mainly washed after every use\cite{klepp2016s}, while a jacket will be cleaned once or twice a year\cite{mccann2009smart, rotzler2021washability}.
\definecolor{lightgray}{rgb}{0.83, 0.83, 0.83}
\begin{table*}
\begin{center}
\begin{NiceTabular}{cccccccc}
\firsthline
\multirow{2}{*}{Structure}&\multicolumn{7}{c}{Properties}\\
\cline{2-8}
(Substrate)[Reference]&\thead{R$_{ext}$\\(A/W)}&\thead{Spectral range\\(nm)}&Rise/Fall time&\thead{D$^{*}$\\(Jones)}&\thead{Bending\\cycle}&\thead{Washing\\cycle}& \\
\hline
\thead{SLG/PVK\\(Fibre) [THIS WORK]}&\thead{2.2$\times$10$^{4}$\\ @1V 488nm}&480-870 & 5/35ms & 10$^{13}$ &\thead{4mm\\100 cycles}&\thead{AATCC\\30 cycles}& \Block[fill=lightgray,borders={left,right}]{10-1}<\rotate>{Flexible PDs} \\
\thead{CuZnS/TiO$_{2}$\\(Fibre) \cite{xu2018real}}&\thead{640\\@3V 320-400nm}&260-420& 200/200ms & - & \thead{60$^{0}$\\200 cycles}&-&  \\
\thead{Carbon fibre/ZnO-CdS\\(Fibre) \cite{zhang2013piezo}}&\thead{10$^{5}$\\@2V 372nm}&372-548& $>$1s & - &\thead{-0.38\%strain\\-}&-&  \\
\thead{P3HT/ZNO\\(Fibre) \cite{du2022piezo}}&\thead{156$\mu$\\@0V 365nm}&365-880& 40/40ms & 0.74x10$^{9}$ &\thead{1.96\% strain\\-}&-&  \\
\thead{Carbon fibre/PVK\\(Fibre) \cite{sun2018ultrahigh}}&\thead{0.563\\@0V 800nm}&350-1050& 200/200ms & 2.15x10$^{13}$ &\thead{90$^{0}$\\60 cycles}& -&  \\
\thead{PVK\\(PET) \cite{hu2014high}}&\thead{3.49\\@3V 365nm}&365-780& 0.14/2.9s & - &\thead{ND\\120 cycles}& - &  \\
\thead{PVK/PDPP3T\\(PET) \cite{chen2016flexible}}&
\thead{154x10$^{-3}$\\@1V 835nm}&365-835& - & 8.8×10$^{10}$ &\thead{7mm\\1000 cycles}& - &  \\
\thead{SLG/MoS$_{2}$ (1L)\\(PET) \cite{de2016high}} &\thead{45.5\\@1V 642nm}&-& - & - &\thead{14mm\\30 cycles} &- &  \\
\thead{InP\\(PET) \cite{yang2010large}} &\thead{0.12\\@-0.3V 533nm}&533-980& - & - &\thead{38.1mm\\-}& - &  \\
\thead{SLG/PVK\\(PA) \cite{dang2016methylammonium}}&\thead{115\\@1V 515nm}&-& 5.3s & 3x10$^{12}$ &\thead{12mm\\3000 cycles}&- &  \\
\hline
\thead{SLG/CsPbBr$_3$\\(Fibre(facet)) \cite{Chen:17}}&\thead{2$\times$10$^{4}$\\ @0.2V 400nm}&400-520 & 24.2s & 8.6x10$^{10}$ & NA & NA &  \Block[fill=lightgray,borders={left,right}]{8-1}<\rotate>{Rigid PDs} \\
\thead{MoS$_{2}$ (1L)\\(Fibre(facet)) \cite{chen2017towards}} &\thead{0.6\\@4V 400nm}&-& 7.1/3.5s & - & NA & NA &  \\
\thead{PVK\\(Glass/ITO) \cite{saidaminov2016perovskite}}&\thead{27\\@5V 830nm}&400-830&30/20$\mu$s & 10$^{13}$ & NA&NA&   \\
\thead{SLG\\(Si/SiO$_{2}$) \cite{mueller2010graphene}}&\thead{6.1$\times$10$^{-3}$\\@0.4V 1150nm}&300-6000& 16GHz & - & NA &NA&   \\
\thead{SLG/PVK\\(Si/SiO$_{2}$) \cite{lee2015high}}&\thead{180\\@0.1V 520nm}&400-980& 87/540ms & 10$^{9}$ & NA &NA&   \\
\thead{SLG/PVK\\(Si/SiO$_{2}$) \cite{xie2017ultrasensitive}}&\thead{10$^{9}$\\@0.5V 598nm}&350-1100& 4.5/57.5s & 10$^{14}$ & NA &NA&   \\
\thead{SLG/QDs\\(Si/SiO$_{2}$) \cite{konstantatos2012hybrid}}&\thead{10$^{7}$\\@5V 532nm}&532-1600& 10/100ms & 7x10$^{13}$ & NA &NA&   \\
\thead{SLG/MoS$_{2}$\\(Si/SiO$_{2}$) \cite{zhang2014ultrahigh}}&\thead{10$^{7}$\\@1V 532nm}&290-680& - & - & NA &NA&   \\
\hline
\thead{Conductive microfluidized\\graphite-coated textile \cite{afroj2020highly}}& NA & NA & NA & NA &\thead{180$^{0}$\\150 cycles}&\thead{BS EN ISO 105 C06 A1S\\ 10 cycles}&  \Block[fill=lightgray,borders={left,right}]{3-1}<\rotate>{Flexible Non PDs}\\
\thead{Polymer\\solar cells \cite{jinno2017stretchable}}& NA & NA & NA & NA &\thead{52\% compression\\20 cycles}&\thead{Submersed in water\\20 cycles}&  \\
\thead{Textile-based\\field-effect transistors \cite{carey2017fully}}& NA & NA & NA & NA &\thead{4mm\\-} &\thead{Tumble-washed\\20 cycles}&  \\
\lasthline
\end{NiceTabular}
\caption{Comparison of KPIs for PDs on rigid and flexible substrates. PA stands for polyamide. NA means not applicable. ND means not defined. The last three rows are neither PDs nor fibre-based structures. Ref.\cite{afroj2020highly} reported 3.5 times higher resistance ($\sim$700Ω/cm) after 10 washing cycles. Ref.\cite{afroj2020highly} used 10 home laundry washing cycles following a British Standard (BS EN ISO 105 C06 A1S). Ref.\cite{jinno2017stretchable} reported$\sim$80\% of the initial power conversion efficiency after water immersion. Ref.\cite{carey2017fully} reported a$\sim$50\% drop in drain current after 20 washing cycles. Our PDs maintain up to$\sim$72\% I$_{ph}$ after 30 AATCC cycles}
\label{table1}
\end{center}
\end{table*}

Bendability can be achieved by integrating rigid components (e.g. Si-based\cite{gupta2018ultra}) within flexible systems\cite{rein2018diode}, including textiles\cite{gao2016fully}. However, device rigidity (i.e. intolerance to various forms of mechanical stress caused by body motions\cite{polat2019flexible}), and washability issues\cite{stoppa2014wearable} are bottlenecks towards commercialization\cite{zeng2014fiber}. Additionally, wafer-based manufacturing processes are generally optimized for planar fabrication\cite{tilli2020handbook} and substrates\cite{sedra2010microelectronic}.

Fibres are an ideal platform for wearable electronics due to their deformability\cite{zeng2014fiber, wang2020flexible} and ease of integration into clothes\cite{zeng2014fiber, wang2020flexible}. They can be used to cover various geometries to make functional surfaces for wearable electronics\cite{son2014multifunctional}. To achieve this, fibre-based conductors, semiconductors and insulators with controlled geometries and interfaces are required. These components could be used to integrate electronics into a textile during weaving\cite{rein2018diode}, for applications such as light-harvesting\cite{li2015wearable,dong2019high}, light-generation\cite{kwon2017weavable}, and photodetection\cite{cai2019materials}. The integration of photodetectors (PDs) into clothes would enable features such as photoplethysmography\cite{polat2019flexible} (i.e. analysing the skin's optical absorption/reflectance to track changes in blood vessels' volume for cardiac pulse measurement\cite{allen2007photoplethysmography}), and hazardous light detection (for skin cancer prevention\cite{xu2018real}).

A number of fibre-based PDs were reported for wearable applications\cite{Chen:17,chen2017towards,zhang2013piezo,sun2018ultrahigh,xu2018real}. These can be compared via a number of key performance indicators (KPIs) including (i) external responsivity\cite{sze2006physics}:
\begin{equation}
R_{ext} = \frac{\triangle {I_{ph}}hc}{q.P_{opt}.\lambda.A_{PD}/A_0}
\end{equation}
where I$_{ph}$ is the photocurrent (I$_{ph}$=I$_{light}$-I$_{dark}$) the difference between current under illumination and in dark, $h$ is the Planck' constant, $c$ is the speed of light, $q$ is the electron charge, $P_{opt}$ is the impinging optical power, $\lambda$ is the wavelength of the incident laser, $A_{PD}$ and $A_{0}$ are the PD area and the laser spot size, respectively. $A_{PD}/A_0$ is a scaling factor taking into account that only a fraction of optical power impinges on the PD, (ii) response time ($\tau_{res}$), i.e. the lifetime of the photogenerated charges in the PD light absorbing layer\cite{koppens2014photodetectors}, (iii) operating voltage\cite{saleh2019fundamentals, koppens2014photodetectors}, (iv) R$_{ext}$ or I$_{ph}$ under mechanical distortion (e.g. bending)\cite{cai2019materials}, and (v) washability, defined as the ability to withstand exposure to water and cleaning processes without being damaged or malfunctioning\cite{cai2019materials}.

State-of-the-art wearable fibre-based PDs have $\tau_{res}\sim$40ms\cite{du2022piezo} and R$_{ext}\sim$156$\mu$A/W\cite{du2022piezo} at 365nm. Under bending strain$\sim$1.96\%, Ref.\cite{du2022piezo} reported that R$_{ext}$ was enhanced$\sim$80\% by the piezo-phototronic effect (a three-way coupling of piezoelectric, semiconductor and photonic properties\cite{du2022piezo}) compared to no strain. Refs.\cite{sun2018ultrahigh, xu2018real} measured $\tau_{res}\sim200ms$ and R$_{ext}\sim$1.48A/W\cite{sun2018ultrahigh} at 800nm and$\sim$640A/W\cite{xu2018real} under UV-A (320-400nm) at 0.5 and 3V, respectively. Bending to 90$\degree$ up to 60 cycles\cite{sun2018ultrahigh} and 60$\degree$ for 200 cycles\cite{xu2018real} resulted in$\sim$8\% and$\sim$14\% decrease of I$_{ph}$, respectively\cite{xu2018real}, see Table 1. However, Refs.\cite{du2022piezo, sun2018ultrahigh, xu2018real} did not discuss washability. Refs.\cite{afroj2020highly,jinno2017stretchable} addressed washability in wearable e-textiles, but did not specifically focus on the washability of PD electronics. Ref.\cite{afroj2020highly} reported microfluidized graphite (20\% $<$10nm thickness after 20 cycles of microfluidization) based poly-cotton fabric\cite{afroj2020highly}, with a polyurethane-based encapsulant to fabricate graphene-coated textile. They measured 3.5 times higher resistance ($\sim$700$\Omega$/cm) after 10 washing cycles\cite{afroj2020highly}. Ref.\cite{jinno2017stretchable} showed washable polymer solar cells made of donor-acceptor polymer with quaterthiophene and naphtho[1,2-$c$:5,6-$c^{'}$]bis[1,2,5]thiadiazole (NTz) (PNTz4T) and [6,6]-phenyl
C$_{71}$-butyric acid methyl ester (PC$_{71}$BM). Ref.\cite{jinno2017stretchable} demonstrated that flexible photovoltaic modules encapsulated with$\sim$1$\mu$m parylene and$\sim$500$\mu$m acrylic elastomers maintained$\sim$80\% of the initial power conversion efficiency (i.e. the percentage of solar energy shining converted into usable electricity\cite{saleh2019fundamentals}) after being submersed in water for 100mins\cite{jinno2017stretchable}. Ref.\cite{chen2017towards} prepared PDs on the end-face of an optical fibre using monolayer (1L) MoS$_2$ prepared by chemical vapor deposition (CVD). This was wet transferred and dip coated onto the fibre facet (end-face)\cite{chen2017towards}. Ref.\cite{chen2017towards} reported R$_{ext}\sim$0.6A/W at 4V and 400nm, with rise time and fall times (i.e. the time for the signal to rise from 10\% to 90\%\cite{saleh2019fundamentals} or fall from 90\% to 10\% of the final value\cite{saleh2019fundamentals})$\sim$7.1s and$\sim$3.5s, respectively\cite{chen2017towards}. However, PD fabrication on fibres' facets is not practical for wearable applications, since fibres' facets can be covered and hindered during weaving\cite{wang2020application}, thus not being able to receive light\cite{cai2019materials}. PDs built parallel to the longitudinal axis of fibres are desired\cite{cai2019materials}, so that the exposed portion receives light, and the covered conductive terminals transmit the signals. Ref.\cite{zhang2013piezo} prepared PDs made of ZnO-CdS double-shell nanowire arrays on carbon fibres\cite{zhang2013piezo}, with R$_{ext}\sim10^{5}$A/W at 372nm at 2V, $\tau_{res}\sim>$1s\cite{zhang2013piezo} with operating wavelength from 372 to 548nm\cite{zhang2013piezo}. Ref.\cite{sun2018ultrahigh} reported fibre-shaped PDs based on CH${_3}$NH$_{3}$PbI$_{3}$ perovskite(PVK)/TiO$_{2}$/carbon fibres and Cu wires, achieving $\tau_{res}\sim$200ms, with R$_{ext}\sim$ 1.48A/W under 0.5V at 800nm and 0.563A/W at a 0V bias at 800nm. Ref.\cite{xu2018real} demonstrated fibre-shaped CuZnS/TiO$_{2}$ nanotubes PDs with Ti wires as inner electrodes and wrapped carbon nanotube around the wire as the outer electrode\cite{xu2018real}, with R$_{ext}\sim$640A/W under UV-A (320-400nm) at 3V, $\tau_{res}\sim$200ms and operating wavelength$\sim$260-420nm, limited by the absorption of TiO$_{2}$ nanotubes\cite{xu2018real}. However, none of these reported washability tests, see Table 1 for a summary.

Graphene and related materials (GRMs) are ideal for a variety of optoelectronics applications\cite{ferrari2015science}, such as PDs\cite{koppens2014photodetectors}, solar cells\cite{bonaccorso2010graphene, yoon2017superflexible, robaeys2014enhanced}, and modulators\cite{sorianello2018graphene,romagnoli2018graphene}. Wafer-scale growth, transfer and fabrication of CVD SLG have improved over the years\cite{tyagi2022ultra, giambra2021wafer, backes2020production}, with carrier mobility, $\mu$, reaching$\sim$18500$cm^{2}V^{-1}s^{-1}$ at room temperature on a rigid substrate such as Si/SiO$_{2}$\cite{lin2019towards}. CVD SLG without encapsulation on ethylene vinyl acetate (EVA)/polyethylene terephthalate (PET) was reported with $\mu$ up to$\sim$7000$cm^{2}V^{-1}s^{-1}$\cite{khan2022high}. However, the absence of a gain mechanism that can generate multiple charge carriers from one incident photon, resulted in limited R$_{ext}\sim$0.01A/W for SLG-only PDs\cite{xia2009ultrafast}.

PVKs are promising photoactive materials, due to their long($>1\mu$m\cite{stranks2013electron}) charge-carrier diffusion lengths\cite{stranks2013electron}, strong ($\sim$10$^{5}cm^{-1}$\cite{de2014organometallic}) absorption coefficients\cite{de2014organometallic}, and bandgap tunability (from 1.1eV\cite{jeon2015compositional} to 2.5eV\cite{jeon2015compositional}) via chemical composition control\cite{jeon2015compositional}. PVKs could be applied for coloured emitters\cite{xing2018color}, energy harvesting\cite{mcmeekin2016mixed}, and PDs\cite{dou2014solution, li2015ambipolar}. PVKs based PDs have low R$_{ext}\sim$0.4A/W in the UV–visible range (350-900nm)\cite{li2020ultrafast, wang2017highly}, mainly due to low $\mu\sim$40$cm^{2}V^{-1}s^{-1}$\cite{abdi2018maximizing}, since R$_{ext}$ is proportional to $\mu$\cite{sze2006physics}:
\begin{equation}\label{MG}
R_{ext}= \frac{\Delta{I}}{P_{opt}} \frac{\tau_{life} \mu V_{ds}}{L^{2}}
\end{equation}
where $\tau_{life}$ is the response time, V$_{ds}$ is the bias applied between source and drain, and $L$ is the channel length. The term $\frac{\tau_{life}\mu V_{ds}}{L^{2}}$ is called gain\cite{sze2006physics}. By increasing $\mu$, the gain increases, which results in higher R$_{ext}$. Consequently, the combination of solution-processed PVK with SLG having $\mu$ well above PVK's, to enhance gain, is promising for light sensing and flexible applications.

Ref.\cite{Chen:17} prepared PDs based on SLG and CsPbBr$_3$ PVK nanocrystals at the end-face of silica fibres with R$_{ext}\sim2\times10^4$A/W under 400nm illumination. SLG was transferred onto the facet, followed by self-assembly of CsPbBr$_3$ nanocrystals\cite{Chen:17}. However, the operating wavelength was$\sim$400-520nm, limited by the CsPbBr$_3$ nanocrystals absorptions\cite{Chen:17}, with $\tau_{res}\sim$24.2s, Table 1.

Here, we report a wet transfer approach to roll CVD SLG around silica fibres, to be used as a gate for flexible PDs. This is followed by the deposition of Al$_2$O$_3$ and parylene C as dielectrics. Another rolled SLG is used as channel with $\mu\sim$1300$cm^{2}V^{-1}s^{-1}$ at RT, before depositing $(Cs_{0.06}FA_{0.79}MA_{0.15})Pb(I_{0.85}Br_{0.15})_{3}$, where MA=methylammonium, $CH_{3}NH^{3+}$ and FA=formamidinium, $CH_{3}(NH_{2})^{2+}$ PVK as photoactive layer. Our PVK does not degrade up to 85$\degree$C\cite{saliba2016cesium} and is chemically stable (ions do not migrate to other layers (i.e. SLG)\cite{tennyson2018cesium}. We get R$_{ext}\sim$22,000A/W at 488nm, with rise and fall times$\sim$5ms and $\sim$35ms, respectively, at 1V, outperforming $R_{ext}$ of fibre-based PDs by at least two orders of magnitude\cite{xu2018real, sun2018ultrahigh, chen2017towards} and one order of magnitude in $\tau_{res}$\cite{xu2018real, sun2018ultrahigh, Chen:17}, Table 1. Bending and washing tests are also performed. After 100 bending cycles at 4mm bending radius I$_{ph}$ is$\sim$80\% of the non-bent one. After 20 and 30 washing cycles according to the AATCC standard\cite{AATCC} we get$\sim$6\% and$\sim$28\% degradation of I$_{ph}$. The washability of our fibre-based PDs goes beyond the typical 20 cycles reported for textile-based field-effect transistors (FETs)\cite{carey2017fully}. Ref.\cite{carey2017fully} had a$\sim$50\% drop in drain current after 20 cycles\cite{carey2017fully}), while our PDs maintain up to 72\% I$_{ph}$ after 30 cycles, compared to unwashed fibre-based PDs, making them promising for wearable applications.
\section{\label{sec:level2}Results and discussion}
\begin{figure*}
\centerline{\includegraphics[width=180mm]{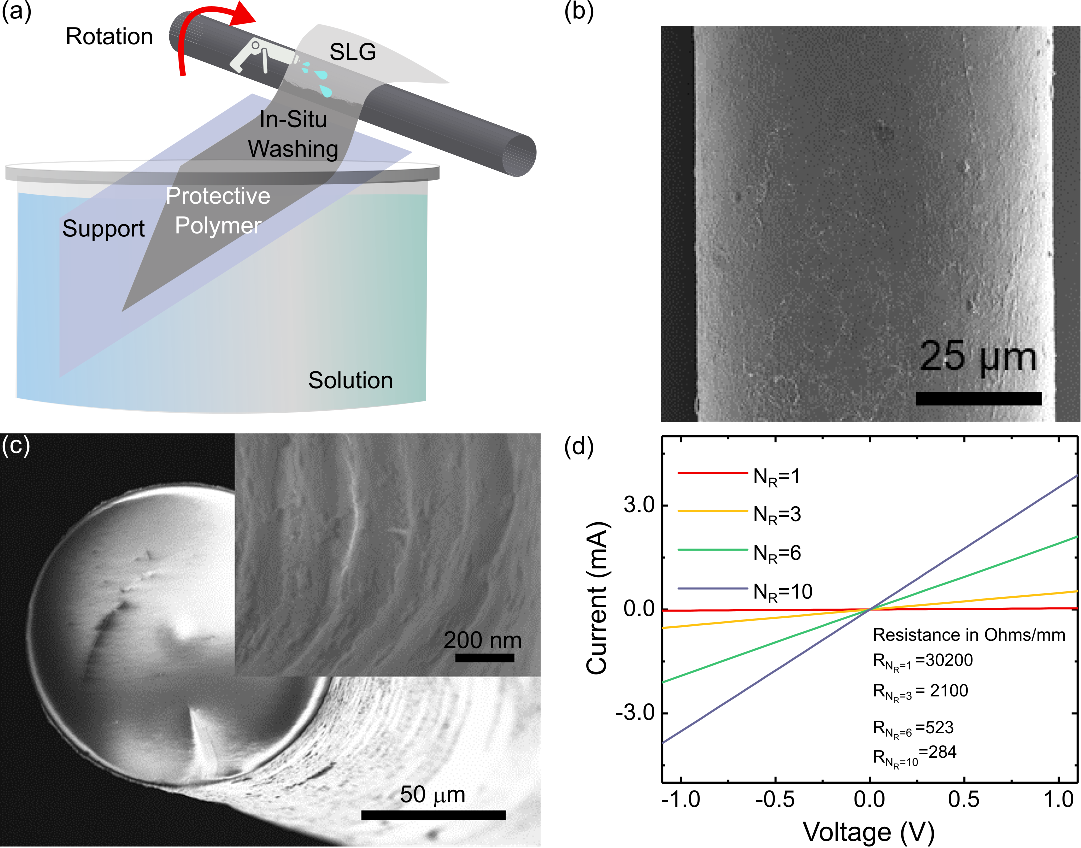}}
\caption{(a) Scheme of the transfer process for CVD SLG around fibres. (b) Top view SEM images for N$_R$=6. (c) Cross-section SEM image for N$_R$=6. (d) The current vs voltage of fibres as a function of N$_R$.}
\label{fig1}
\end{figure*}
Fig.\ref{fig1} plots a scheme of the transfer process of CVD SLG around fibres. CVD SLG is prepared on a $35\mu{m}$ Cu foil as for Ref.\cite{bae2010roll}. The Cu foil is annealed up to 1000$^{\circ}$C for 30mins under H$_{2}$ at 20sccm. Then, a gas mixture of CH$_4$ (5sccm) and H$_2$ (20sccm) is added to initiate growth at 1000$^{\circ}$C for 30mins. Finally, the sample is cooled in vacuum (1mTorr) to RT. In order to transfer SLG, poly(methyl methacrylate) (PMMA) dissolved in acetone is spin coated on SLG/Cu at 4000rpm for 40s, followed by oxygen etching of SLG on the Cu backside using an RIE-NanoEtch (3W, 30s)\cite{bonaccorso2012production, bae2010roll}. Cu/SLG/PMMA is then transferred to an aqueous ammonium persulfate (APS) solution ((NH$_4$)$_2$S$_2$O$_8$, 0.5 mol/L) for Cu dissolution\cite{bonaccorso2012production, bae2010roll}. The PMMA membrane and CVD SLG are then cleaned of APS residuals by dipping in DI water for 20mins. This is followed by the rolling process. First, the fibre is placed on a supporting film (e.g. poly(ethylene terephthalate) (PET)) and the edge of the floating SLG/PMMA is fished onto the fibre, while the rest of the SLG/PMMA stays on the supporting film. Then, the fibre is rotated clockwise at$\sim$10rpm to roll SLG/PMMA. The PMMA layer is then removed by immersing the coated fibre in acetone for 15mins, followed by rinsing with isopropanol for 10mins and drying in N$_2$. Fig.\ref{fig1}b is the top view of 6 SLG layers coated on fibre glass, taken by scanning electron microscopy (SEM, Magellan 400L). Fig.\ref{fig1}c shows the cross-section.

Fig.\ref{fig1}d plots the conductivity (measured via a Keithley source meter at the two ends of the channel layer) of different coated fibres depending on the number of rolled SLG, N$_R$, for a channel length$\sim$1mm. The linear relation between current and voltage indicates Ohmic conductivity. Given that increasing N$_R$ reduces sheet resistance\cite{bonaccorso2010graphene} due to increased levels of conductive paths\cite{bakhshaee2022multilayer}, N$_R$=6 is chosen as a gate electrode. N$_R$=10 leads to 284$\Omega$/mm, however the roughness caused by PMMA residuals between layers could result in a short circuit when the dielectric layer is positioned between gate and channel layer. Thus, we use N$_R$=6 to fabricate conductive fibres with a uniform surface coverage of SLG as for Figs.\ref{fig1}b,c.

\begin{figure}
\centerline{\includegraphics[width=90mm]{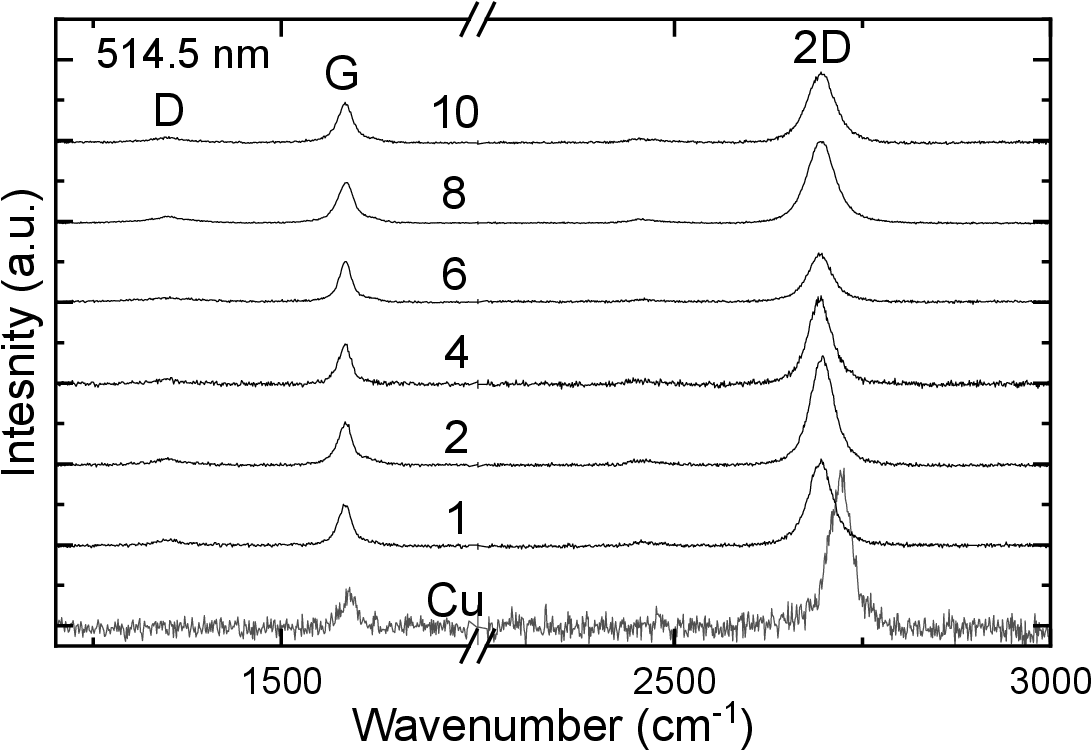}}
\caption{514.5nm Raman spectra of CVD SLG on Cu and for N$_R$=1, 2, 4, 6, 8, 10, normalized to have the same I(G).}
\label{fig2_1}
\end{figure}

The presence and quality of SLG are further studied by Raman spectroscopy at 514.5nm using a Renishaw InVia with a 50$\times$ objective and $<$0.1mW power on the sample. Fig.\ref{fig2_1} plots the Raman spectrum of the film as grown on Cu. The 2D peak is a single Lorentzian with FWHM(2D)$\sim$34cm$^{-1}$, signature of SLG\cite{ferrari2006raman}. The position of the G peak, Pos(G), is$\sim$1591cm$^{-1}$, with FWHM(G)$\sim$22cm$^{-1}$. The 2D peak position, Pos(2D), is$\sim$2720cm$^{-1}$, with FWHM(2D)$\sim$34cm$^{-1}$, while the 2D to G peak intensity and area ratios, I(2D)/I(G) and A(2D)/A(G), are$\sim$4.4 and$\sim$6.7. No D peak is observed, indicating negligible defects\cite{ferrari2013raman}. Fig.\ref{fig2_1} plots Raman spectra of CVD SLG for N$_{R}$=1, 2, 4, 6, 8, 10. The 2D line shape is independent of N$_R$, suggesting no interaction between transferred SLGs\cite{ferrari2013raman, ferrari2007raman}, Fig.\ref{fig2_1}.

\begin{figure}
\centerline{\includegraphics[width=90mm]{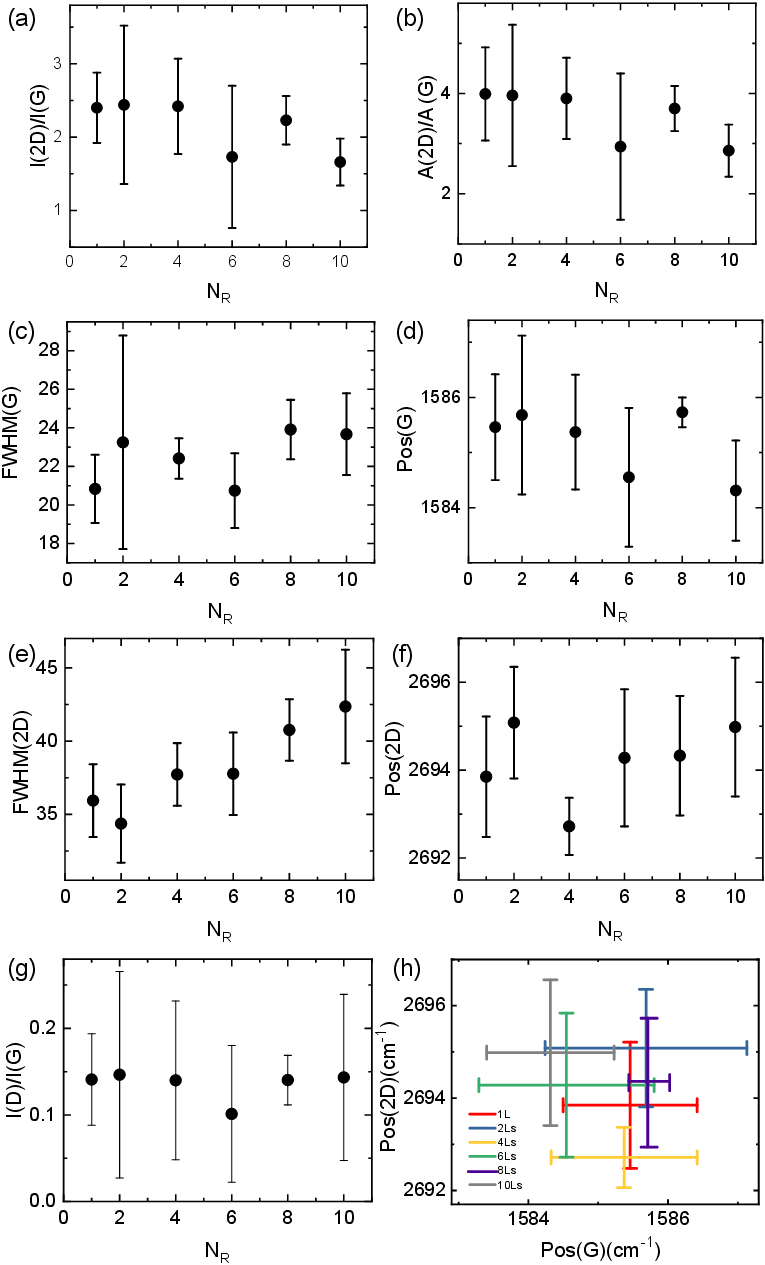}}
\caption{514.5nm Raman spectra for N$_R$=1, 2, 4, 6, 8, 10. For each N$_R$, 10 points are collected along the fibre. (a-g) Average and standard deviations of (a) I(2D)/I(G), (b) A(2D)/A(G), (c) FWHM(G), (d) Pos(G), (e) FWHM(2D), (f) Pos(2D), (g) I(D)/I(G). (h) Pos(2D) as a function of Pos(G).}
\label{fig2_2}
\end{figure}

Fig.\ref{fig2_2} plots Pos(G), FWHM(G), Pos(2D), FWHM(2D), I(2D)/I(G), A(2D)/A(G), I(D)/I(G) as a function of N$_R$. For each N$_R$, 10 points are collected along the fibre. I(2D)/I(G) changes from$\sim$2.4 for N$_R$=1 to$\sim$1.7 for N$_R$=10, Fig.\ref{fig2_2}a. A(2D)/A(G) changes from$\sim$4 for N$_R$=1 to $\sim$2.9 for N$_R$=10, Fig.\ref{fig2_2}b. The decrease in I(2D)/I(G) and A(2D)/A(G) indicates doping increases from E$_F\sim$225meV for N$_R$=1 to$\sim$320meV for N$_R$=10\cite{basko2009electron, das2008monitoring}. This is further supported by the decrease of Pos(G) as function of N$_R$ from$\sim$1585.4cm$^{-1}$ for N$_R$=1 to 1584.3cm$^{-1}$ for N$_R$=10, Fig.\ref{fig2_2}d. From I(D)/I(G), Fig.\ref{fig2_2}g, the defect density changes from$\sim$5.74$\times$10$^{10}$cm$^{-2}$ for N$_R$=1 to 7.1$\times$10$^{10}$cm$^{-2}$ for N$_R$=10\cite{canccado2011quantifying, bruna2014doping} for excitation energy 2.41eV and E$_F\sim$225meV and 320meV, respectively, Fig.\ref{fig2_2}g. FWHM(2D,G) are broadened as N$_R$ increases due to defects and inhomogeneities of rolled CVD graphene, Figs.\ref{fig2_2}c,e.

For N$_R$=6, Pos(G)$\sim$1584cm$^{-1}$, FWHM(G)$\sim$21cm$^{-1}$, Pos(2D)$\sim$2694cm$^{-1}$, FWHM(2D)$\sim$38cm$^{-1}$, I(2D)/I(G)$\sim$1.7 A(2D)/A(G)$\sim$2.9, indicating p doping with E$_F\sim$310meV by  taking  into  account  the  dielectric constant$\sim$3.9 of the silica fibre\cite{el2012fundamentals}. I(D)/I(G)$\sim$0.1 corresponds to a defect density$\sim$4.9$\times$10$^{10}$cm$^{-2}$\cite{canccado2011quantifying, bruna2014doping}. Pos(G,2D) are affected by the presence of strain. Biaxial strain can be differentiated from uniaxial from the absence of G-peak splitting with increasing strain\cite{mohiuddin2009uniaxial}, however at low ($\lessapprox$0.5\%) strain the splitting cannot be resolved\cite{mohiuddin2009uniaxial, yoon2011strain}. For uniaxial(biaxial) strain, Pos(G) shifts by $\Delta$Pos(G)/$\Delta$ $\epsilon_{strain}\sim$23(60)cm$^{-1}$/\%\cite{mohiuddin2009uniaxial, yoon2011strain}. Pos(G) also depends on doping\cite{das2008monitoring, basko2009electron}. However, Pos(2D) and Pos(G) are not correlated, Fig.\ref{fig2_2}h. Thus, the spread of Pos(G) and Pos(2D) is mostly due to doping. Table 2 summarizes the Raman fit parameters and resulting E$_{F}$, doping type, strain, defect density.
\begin{figure}
\centerline{\includegraphics[width=90mm]{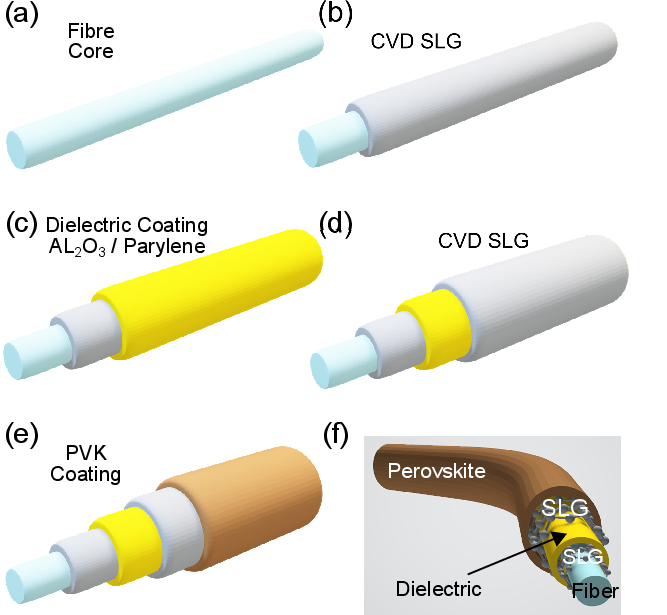}}
\caption{Fibre-based SLG/PVK PDs fabrication process. (a) Silica fibre substrate. (b) N$_R$=6 around the fibre as gate electrode. (c) Gate dielectric layer composed of Al$_{2}$O$_{3}$ and parylene C. (d) Rolled N$_R$=6 channel. (e) Spin-coated PVK on rolled CVD SLG channel. (f) Assembled fibre-based SLG/PVK PD.}
\label{fig3}
\end{figure}
\begin{figure*}
\centerline{\includegraphics[width=180mm]{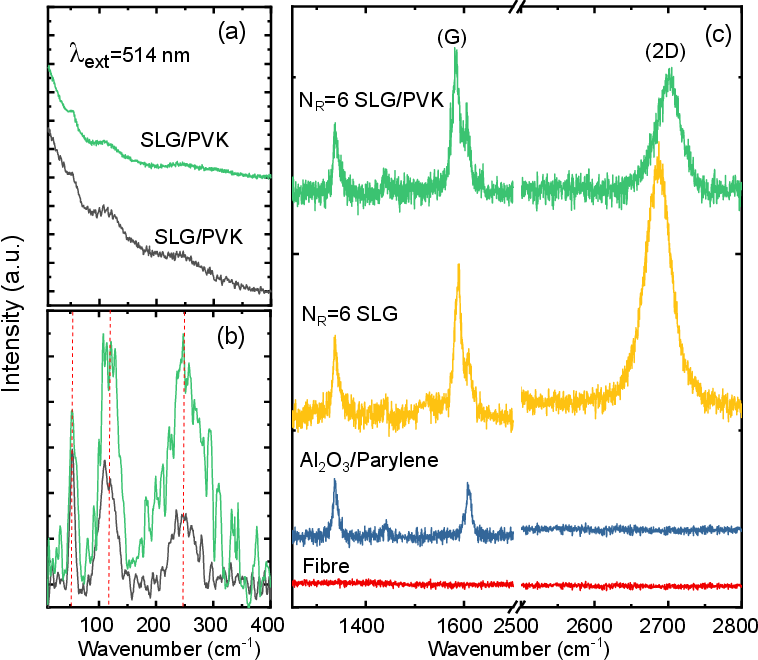}}
\caption{(a) Raman spectra of SLG/PVK on fibre (green) and SLG/PVK on Si/SiO${_2}$ (gray). (b) Backgrounds subtracted spectra from (a). (c) Raman spectra of fibre (red), fibre/dielectric (Al$_{2}$O$_{3}$ and parylene) (blue), fibre/dielectric/N$_{R}$=6 (orange), fibre/dielectric/N$_{R}$=6/PVK (green). The PL background of PVK is subtracted for the green spectrum. All spectra for 514.5nm.}
\label{fig4}
\end{figure*}

Fig.\ref{fig3} is a schematic drawing of our fibre-based PDs. The fabrication process starts by wiping the silica fibre via acetone and isopropanol, Fig.\ref{fig3}a. Then, CVD SLG is rolled 6 times around the fibre (reaching a conductivity$\sim$500$\Omega$/mm to form a gate electrode, Fig.\ref{fig3}b). Then, 150nm Al$_2$O$_3$ and 200nm parylene C are deposited as gate-dielectrics, Fig.\ref{fig3}c. Al$_2$O$_3$ is commonly used as a gate dielectric in graphene field-effect transistors\cite{fallahazad2012scaling} because of its dielectric constant ($k\sim$9)\cite{gupta2010copper, kim2009realization}, compatibility with graphene\cite{kim2009realization} and breakdown voltage (up to 10 MV/cm)\cite{gupta2010copper}. Parylene is also commonly employed as dielectric, pin-hole free\cite{stassen2004influence, sabri2009graphene}, passivation layer\cite{chua2005general} in flexible electronics\cite{yin2018solution, marszalek2017parylene}. Al$_2$O$_3$ coating is performed by atomic layer deposition (ALD, Cambridge Nanotech Savannah), to achieve uniform coating and film thickness control (down to the nm range)\cite{puurunen2005surface}. We deposit a 150nm-thick Al$_2$O$_3$ dielectric layer at 150$^{\circ}$C at a base pressure$\sim$0.5mbar using trimethylaluminum (TMA purity $>$98\%, Strem Chemicals 93-1360) as precursor and H$_{2}$O as oxidant. Oxidant and precursors are alternately introduced into the chamber in a 20sccm flow of N$_{2}$ carrier gas. If Al$_2$O$_3$ alone is used, we get dielectric failure because of crack formation, due to the mechanical stress/strain involved in the fabrication process of our PDs. 200nm parylene is applied as follows. A parylene dimer is vaporized at 80$^{\circ}$C, then, in a separate chamber, it is pyrolysed into monomers at 690$^{\circ}$C. The fibre is held at room temperature (RT), so that the parylene polymerizes on contact with the surface. The physical vapor deposition of parylene provides coverage of all accessible surfaces\cite{slinker2007direct}. This improves dielectric robustness against deformations during manufacturing of fibre-based PDs, reducing the risk of short circuits. The sole use of parylene C, however, creates inconsistent dielectric properties: thin layers ($\sim$500nm) result in gate leakage at low ($\sim$0.2MV/m) electric fields, while exceeding 500nm compromises the channel modulation due to a decrease in gate dielectric capacitance\cite{kahouli2012effect}.
\begin{figure*}
\centerline{\includegraphics[width=170mm]{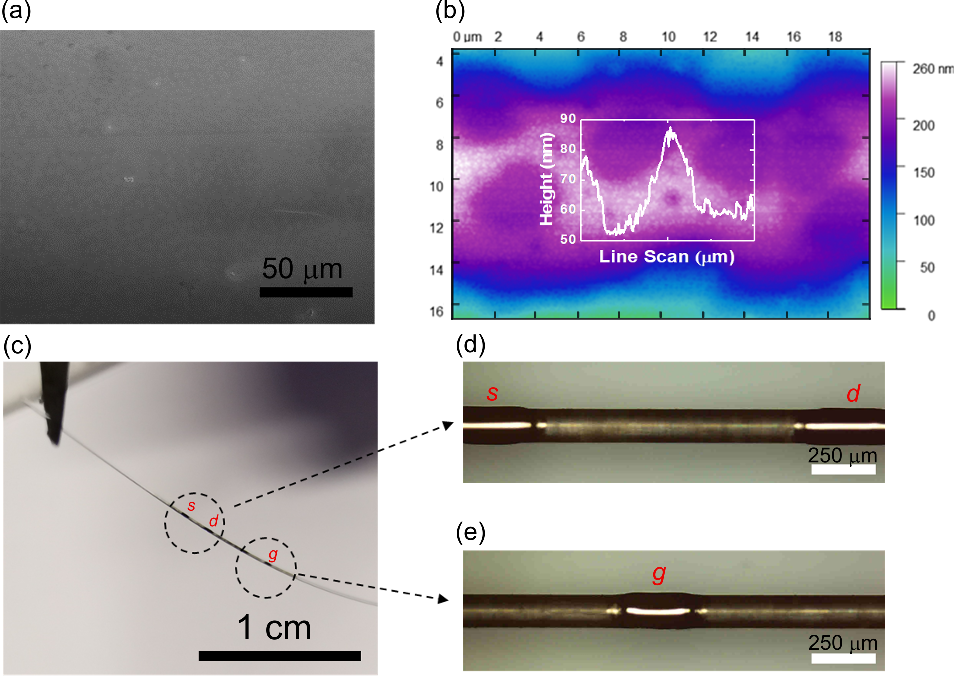}}
\caption{(a) SEM and (b) AFM image of PVK film on N$_R$=6, inset image shows the roughness of spin-coated PVK film. (c-e) Optical microscopy images of final device. (d,e) Source-drain and gate electrodes, respectively, printed using Ag ink. s, d, g indicate source, drain, gate, respectively}
\label{fig5}
\end{figure*}
\begin{table*}
\begin{center}
\begin{tabular}{cccccccccccc}
\firsthline
\multicolumn{12}{c}{}\\
Sample& \thead {Pos(G)\\cm$^{-1}$}&\thead {FWHM(G)\\cm$^{-1}$}&\thead {Pos(2D)\\cm$^{-1}$}&\thead{FWHM(2D)\\cm$^{-1}$}&A(2D)/A(G)&I(2D)/I(G)&I(D)/I(G)&\thead {E$_{F}$\\meV}&\thead {Doping\\ type}&\thead {Uniaxial\\strain\%} \thead {Biaxial\\strain\%}&\thead {defect density\\$\times$10$^{11}$cm$^{-2}$}\\
\hline
N$_R$=6 &\thead{1584\\$\pm$1}&\thead{21\\$\pm$2}&\thead{2694\\ $\pm$2}&\thead{38\\ $\pm$3}&\thead{2.9\\ $\pm$1.5}&\thead{1.7\\ $\pm$0.9}&\thead{0.10\\ $\pm$0.08}&\thead{310\\$\pm$110}&p&\thead{0.40 \\ $\pm$0.04} \thead{0.15 \\ $\pm$0.02}&\thead {0.49\\$\pm$0.12}\\
\hline
N$_R$=6 &\thead{1587 \\ $\pm$1}&\thead{17 \\ $\pm$2}&\thead{2686 \\ $\pm$3}&\thead{41 \\ $\pm$3}&\thead{4.7 \\ $\pm$1}&\thead{1.9 \\ $\pm$0.3}&\thead{0.49 \\ $\pm$0.1}&\thead{190 \\ $\pm$20} &p&\thead{-0.06\\ $\pm$0.06}\thead{-0.02\\ $\pm$0.02}&\thead{1.82 \\ $\pm$0.39}\\

N$_R$=6/PVK &\thead{1583 \\ $\pm$1}&\thead{17 \\ $\pm$2}&\thead{2701\\ $\pm$3}&\thead{43\\ $\pm$3}&\thead{2.3\\ $\pm$0.6}&\thead{0.9\\ $\pm$0.3}&\thead{0.46\\ $\pm$0.1}&\thead{380\\$\pm$30}&p&\thead{0.60 \\ $\pm$0.03} \thead{0.20\\ $\pm$0.01}&\thead{2.48\\ $\pm$0.62}\\
\lasthline
\end{tabular}
\caption{Raman fit parameters, resulting Fermi energy (E$_{F}$), doping type (n or p), strain, and defect density. The first row relates to rolled graphene layers on top of the fibre (gate), while the last two rows correspond to fibre with a dielectric layer (150nm Al$_{2}$O$_{3}$ and 200nm parylene C) (channel).}
\label{table2}
\end{center}
\end{table*}

Fig.\ref{fig4} plots the Raman spectra of the fibre PD at different fabrication stages. Raman spectra of fibre (red), dielectric layer (150nm Al$_2$O$_3$ and 200nm parylene C) (blue), N$_{R}$=6 on the dielectric layer (green), and coated PVK on N$_{R}$=6 (orange). In the Raman spectrum of Al$_2$O$_3$ and parylene C, Fig.\ref{fig4}c, the peaks$\sim$1337cm$^{-1}$, 1440cm$^{-1}$, 1610cm$^{-1}$, Fig.\ref{fig4}c, are attributed to CH$_{2}$ wagging and twisting vibrations\cite{mathur1973laser, jakabovivc2009preparation}, CH scissoring in CH$_{2}$ and/or C-C skeletal in-plane vibrations of aromatic rings\cite{mathur1973laser, jakabovivc2009preparation}, and CH scissoring in CH$_{2}$ and/or C-C skeletal in-plane vibrations of aromatic rings\cite{mathur1973laser, jakabovivc2009preparation}, respectively. For N$_{R}$=6, Pos(G)$\sim$1587cm$^{-1}$, FWHM(G)$\sim$17cm$^{-1}$, Pos(2D)$\sim$2686cm$^{-1}$, FWHM(2D)$\sim$41cm$^{-1}$, I(2D)/I(G)$\sim$1.9 and A(2D)/A(G)$\sim$4.7. Pos(2D)$\sim$2686cm$^{-1}$ indicates that layers are p-doped\cite{basko2009electron}. From A(2D)/A(G)$\sim$4.7 we estimate E$_F\sim$190meV, corresponding to a carrier concentration$\sim$2.2$\times$10$^{12}$cm$^{-2}$ by taking into account the dielectric constant$\sim$3\cite{mark2004physical} of Parylene and the residual PMMA from transfer of CVD SLG\cite{basko2009electron}. I(D)/I(G)$\sim$0.49 corresponds to a defect density$\sim$1.82$\times$10$^{11}$ cm$^{-2}$\cite{canccado2011quantifying, bruna2014doping} for 2.41eV excitation and E$_F\sim$190meV. The doping level as calculated from A(2D)/A(G) should correspond to Pos(G)$\sim$1586cm$^{-1}$ for unstrained graphene\cite{das2008monitoring}. However, in our experiment Pos(G)$\sim$1587cm$^{-1}$, which implies a contribution from compressive uniaxial (biaxial) strain$\sim$0.06\% (0.02\%)\cite{mohiuddin2009uniaxial}. Table 2 summarizes the Raman fit parameters and the resulting E$_{F}$, doping, strain, and defect density.

Source and drain electrodes are defined (channel length: 1000$\mu$m; diameter: 125$\mu$m) by inkjet printing Ag ink (Sigma-Aldrich, 736465) (resistivity=11$\mu\Omega$cm). Ag ink is also printed on the gate, and the ink is cured on a hot plate at 150$^{\circ}$C for 1h to remove the residual solvent (triethylene glycol monomethyl ether). We use a Fujifilm Dimatix DMP-2800 with a 21$\mu$m diameter nozzle. This approach removes the need for conventional lithography\cite{ng2019printing}, or shadow masks\cite{plummer2009silicon}.

Then, a mixed-cation lead mixed-halide PVK\cite{abdi2018maximizing} is prepared by dissolving PbI$_{2}$ (1.2M), formamidinium iodide (1.11M), methylammonium bromide (0.21M) and PbBr$_{2}$ (0.21M) in a mixture of anhydrous DMF:DMSO (4:1 volume ratio) followed by 5 vol\% CsI stock solution (1.5M in DMSO). The final solution is diluted by adding 50 vol\% DMF:DMSO (4:1 volume ratio). The PVK solutions is spin-coated (Fig.\ref{fig3}e) on the surface-treated rolled SLG surface (10min under UV ozone utilising UV-Ozone Cleaner UVC-1014, in order to facilitate the dispersion of PVK, by making the surface hydrophilic\cite{liu201920}) using a two-step speed program at 2000 and 6000rpm for 10 and 30s, respectively, adding 150$\mu$l of chlorobenzene 30s after the start of the spinning routine\cite{jeon2014solvent}. These different speeds allow us to get more homogeneity with respect to one-step spinning\cite{jeon2014solvent}, and remove solvents during the PVK film formation\cite{jeon2014solvent}. The samples are annealed at 100$^{\circ}$C for 1h to remove the solvents (Dimethylformamide and Dimethyl sulfoxide)\cite{jeon2014solvent}. All processes to make and spin coat PVK are performed in a N$_2$-filled glove box to prevent degradation of PVK triggered by H$_{2}$O\cite{berhe2016organometal}.

Fig.\ref{fig4}a plots the Raman spectra of SLG/PVK on fibre (green) and Si/SiO$_{2}$ (gray) for 514.5nm excitation. These show three bands$\sim$52, 110, 249 $cm^{-1}$. The PL background of PVK can be seen in Fig.\ref{fig4}a and background subtracted spectra are in Fig.\ref{fig4}b. The bands$\sim$52 and 110$cm^{-1}$ are the main modes of PVK\cite{2015raman} assigned to PbBr$_{6}$ deformation\cite{calistru1997identification, dammak2007x} and motion of the organic cation\cite{calistru1997identification, dammak2007x}, respectively. The peak$\sim$249$cm^{-1}$ is assigned to torsional mode of the methylammonium cation, due to PVK exposure to moisture\cite{quarti2013raman, 2015raman}. Moisture could be introduced when the sample is removed from the glove-box, prior to sealing with parylene. After the deposition of PVK on N$_{R}$=6, Fig.\ref{fig3}e, the Raman spectrum of N$_{R}$=6 has Pos(G)$\sim$1583cm$^{-1}$, FWHM(G)$\sim$17cm$^{-1}$, Pos(2D)$\sim$2701cm$^{-1}$, FWHM(2D)$\sim$43cm$^{-1}$, I(2D)/I(G)$\sim$0.9 and A(2D)/A(G)$\sim$2.3. Pos(2D)$\sim$2701cm$^{-1}$ indicates p-doping\cite{basko2009electron}. From A(2D)/A(G)$\sim$2.3 we estimate E$_F\sim$380meV, which corresponds to a carrier concentration$\sim$8.8$\times$10$^{12}$cm$^{-2}$, by taking into account the dielectric constant of Parylene and residual PMMA\cite{basko2009electron}. I(D)/I(G)$\sim$0.46 corresponds to a defect density$\sim$2.5$\times$10$^{11}$ cm$^{-2}$\cite{canccado2011quantifying, bruna2014doping} for 2.41eV excitation and E$_F\sim$380meV. E$_F$ as estimated from A(2D)/A(G) should correspond to Pos(G)$\sim$1596cm$^{-1}$ for unstrained graphene\cite{das2008monitoring}. However, in our experiment Pos(G)$\sim$1583cm$^{-1}$, which implies tensile uniaxial (biaxial) strain$\sim$0.6\%(0.2\%)\cite{mohiuddin2009uniaxial}.
\begin{figure}
\centerline{\includegraphics[width=90mm]{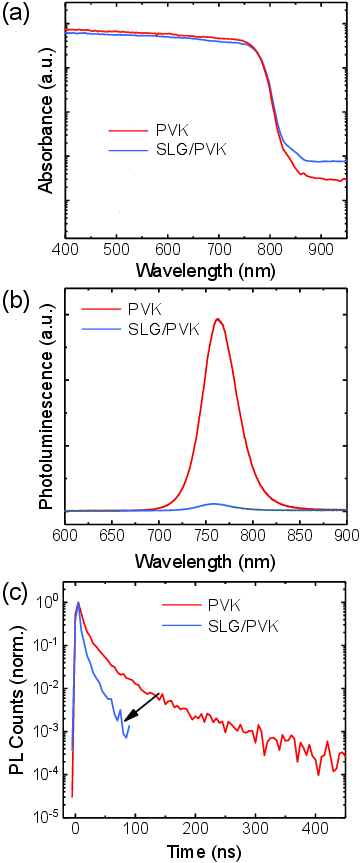}}
\caption{(a) Absorbance of PVK and SLG/PVK. (b) PL of PVK and SLG/PVK for 514.5nm excitation. (c) TRPL spectra of PVK and SLG/PVK for 400nm excitation. The arrow indicates a shorter lifetime.}
\label{fig6}
\end{figure}
\begin{figure*}
\centerline{\includegraphics[width=170mm]{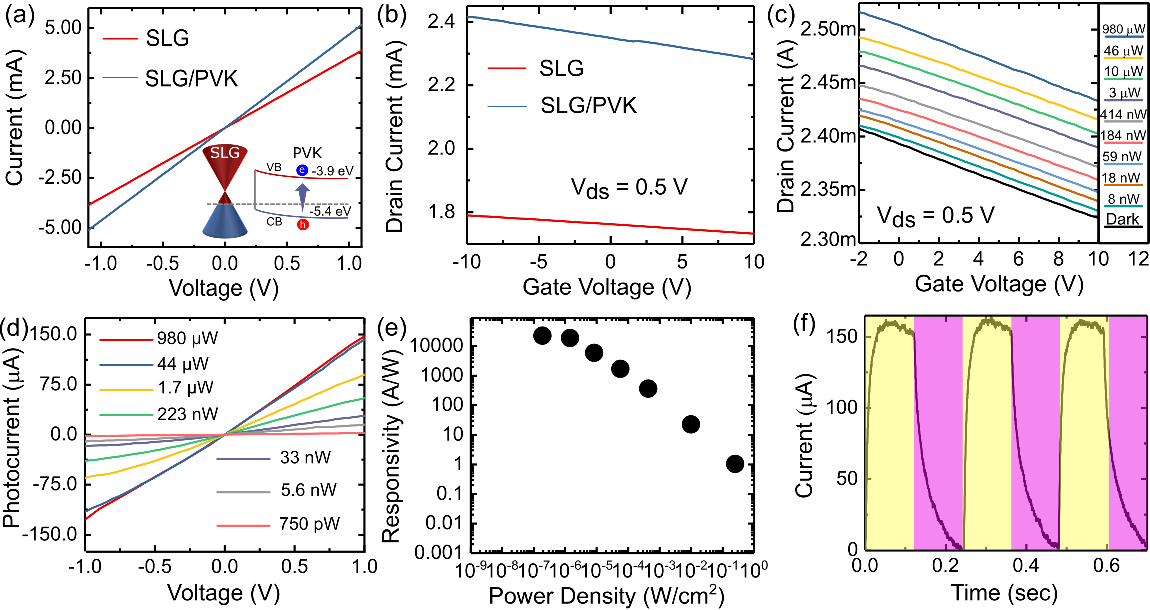}}
\caption{(a) I$_{ds}$ as a function of V$_{ds}$ in the dark before (red) and after (blue) deposition of PVK on N$_R$=6. (b) Transconductance for V$_{ds}$=0.5V for N$_R$=6 in dark. (c) Illumination power-dependent transfer characteristics at 488nm. (d) Photocurrent as a function of bias for different illumination power and V$_g$=0. (e) R$_{ext}$ as a function of optical power density. (f) Temporal photocurrent response for N$_R$=6 under alternating dark (purple) and light (yellow).}
\label{fig8}
\end{figure*}

Fig.\ref{fig5}a is a SEM image of the spin-coated PVK on N$_R$=6. This shows uniform coverage. An average roughness$\sim$20nm is measured with a Bruker Dimension Icon atomic force microscope (AFM) in Fig.\ref{fig5}b. Figs.\ref{fig5}c-e are optical images of the final device.

Fig.\ref{fig6}a plots the absorbance from photo-thermal deflection spectroscopy (PDS) of individual SLG and PVK films on quartz, compared with SLG/PVK. A combination of a Light Support MKII 100 W Xe arc source and a CVI DK240 monochromator is used to produce a modulated monochromated light beam. The probe beam comes from a Qioptiq 670nm fibre-coupled diode laser directed parallel to the PVK film surface. PDS can measure the band-tails of PVK down to$\sim$10$^{-5}$absorbance\cite{pazoki2019characterization}. Fig.\ref{fig6}a shows that PVK absorption increases for SLG/PVK compared to PVK. This indicates the presence of interface states that alter the PVK bandgap\cite{pazoki2019characterization}. The PL spectra of PVK and SLG/PVK are in Fig.\ref{fig6}b. Both show a PL peak$\sim$762nm arising from the PVK band gap\cite{abdi2018maximizing}. The PL intensity (integrated area under PL curve) of SLG/PVK is quenched$\sim$96\% compared to PVK. This can be assigned to charge carrier transfer between PVK and SLG\cite{zhu2014efficiency}, though there is also increased non-radiative recombination due to the increased sub-gap trap density, as shown by the PDS measurements in Fig.\ref{fig6}a. Time-resolved PL (TRPL) spectra are recorded at RT with a gated intensified CCD camera (Andor iStar DH740 CCI-010) connected to an Andor SR303i spectrometer with time resolution$\sim$0.1ns. A Ti:sapphire optical amplifier (1kHz, 90fs pulse width) is used to generate narrow bandwidth photoexcitation (FWHM=10nm at 400nm) via a noncollinear optical parametric amplifier and a pulse fluence of 0.5$\mu$Jcm$^{-2}$. We thus fit Fig.\ref{fig6}c to calculate average lifetimes as a means to compare the amount of charge quenching, but we do not imply any physical meaning to this fit. We use the relation\cite{saleh2019fundamentals}:
\begin{equation}
f(t) = \sum_{i} A_{i} exp(-t/\tau_{i})+B
\end{equation}
where $A_{i}$ is decay amplitude, $\tau_{i}$ is decay time and B is constant. The PL decay times for PVK are $\tau_{1}$=6.7ns (A$_{1}$:0.802) and $\tau_{2}$=32.9ns (A$_{2}$:0.198) with average lifetime ($\Sigma A_{i}\tau_{i)}$)/($\Sigma A_{i}$) of 11.8ns. The PL decay times for SLG/PVK are $\tau_{1}$=2.0ns (A$_{1}$:0.743) and $\tau_{2}$=10.7ns (A$_{2}$:0.257) with average lifetime$\sim$4.3ns, indicating charge transfer between SLG and PVK\cite{li2017high}, because of band energy alignment at the PVK/SLG interface.

We then characterize the photoresponse of the fibre-based PDs. The linear dependence of current vs voltage in Fig.\ref{fig8}a indicates Ohmic contact of CVD SLG and Ag electrodes. By depositing PVK, the current increases in Figs.\ref{fig8}a,b, suggesting h transfer from PVK to SLG, in agreement with the Raman measurement in Fig.\ref{fig4}. A similar h transfer was previously reported for SLG/PVK interfaces\cite{lee2015high, wang2015hybrid}. The field-effect $\mu$ is calculated as\cite{sze2006physics}:
\begin{equation}
\mu = \frac{\triangle I_d.L}{\triangle V_g.C_{ox}.V_{ds}.2\pi r}
\end{equation}
\begin{figure}
\centerline{\includegraphics[width=90mm]{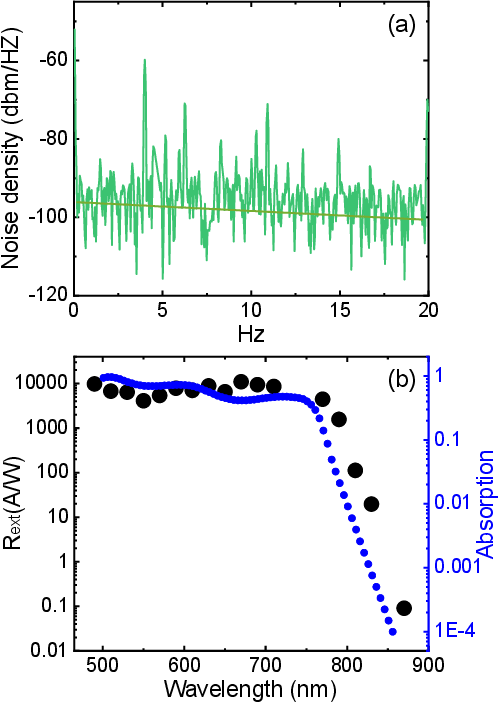}}
\caption{(a) Spectral noise power of fibre PDs. (b) Wavelength dependent R$_{ext}$ for PDs and absorption of fibre PDs from the KK relations as a function of excitation wavelength}
\label{fig9}
\end{figure}

Fig.\ref{fig8}c plots I$_{ds}$ vs V$_{g}$ when the PD is illuminated at 488nm for normal incidence with a collimated laser (Multi-Channel Source-Thorlabs) with P=44$\mu$W to 2nW for V$_{ds}\sim$0.5V. -2$< V_{g}<$+10 is used as a safe range to avoid dielectric breakdown. The gate-source current is kept$<$1nA to avoid leakage. Due to the ultrafast (sub-100fs) charge recombination in SLG\cite{brida2013ultrafast}, the photogenerated carriers do not contribute to the photocurrent. A gate-dependent photocurrent is not observed in our channel, Fig.\ref{fig8}c. This could be explained by the screening of the electric field by the SLG layers underneath\cite{santos2013electric}. Under illumination, light is absorbed by PVK and part of the photogenerated h are transferred from the PVK valence band into the lower energy states in p-doped SLG\cite{lee2015high, wang2015hybrid}, leaving behind the uncompensated charge of photogenerated e\cite{lee2015high, wang2015hybrid}. The latter are trapped in PVK and act as an additional negative gate voltage, applied to the SLG channel, altering the electric field at the SLG/PVK junction\cite{lee2015high, wang2015hybrid}. The photocurrent is\cite{sze2006physics}:
\begin{equation}
\triangle {I_{ph}} = I_{light}-I_{dark}
\end{equation}
where $I_{light}$ is the output current of the device under the incident beam, and $I_{dark}$ is the current measured in dark. The photocurrent of the fibre-based SLG/PVK PDs is in Fig.\ref{fig8}d, for P ranging from 980$\mu$W to 750pW, and V$_{ds}$ from -1 to 1V. The corresponding R$_{ext}$ as a function of power density is in Fig.\ref{fig8}e. For A$_{PD}$ and A$_{0}\sim$0.18$\mu$m$^{2}$ and 0.785$\mu$m$^{2}$, respectively, we get R$_{ext}\sim$22,000A/W for P=750pW and V$_{ds}$=1V. For V$_{ds}>$1, the free carriers drift velocity $\nu_{d}$ in SLG increases lineally until saturation by scattering with optical phonons\cite{lazzeri2005electron}. Therefore, all measurements are done at V$_{ds}\leq$1V to keep the device operation in the linear (Ohmic) regime. A P rise leads to higher photocurrents until saturation (44$\mu$W and 980$\mu$W), due to the increased recombination rate of photo excited carriers\cite{wang2015hybrid}. This R$_{ext}$ is seven orders of magnitude higher than the$\sim$mA/W reported for SLG on flat and rigid substrates (e.g. Si/SiO$_{2}$)\cite{xia2009ultrafast, mueller2010graphene, gan2013chip, pospischil2013cmos}, one order of magnitude higher than reported for PVK on Si/SiO$_{2}$\cite{li2015ambipolar}, and four orders of magnitudes higher than pristine PVK$\sim$3A/W\cite{hu2014high} on flexible PET substrates. Our R$_{ext}$ is two orders of magnitude higher than PVK/SLG hybrid PDs on Si/SiO$_{2}$\cite{lee2015high}, but lower than the highest R$_{ext}\sim$10$^{9}$A/W PDs based on PVK/organic-semiconductor(poly-(3,4-ethylenedioxythiophene):poly(styrenesulfonate))\cite{xie2017ultrasensitive} and lower than hybrid SLG/QDs\cite{konstantatos2012hybrid} and SLG/MoS$_{2}$ PDs\cite{zhang2014ultrahigh}, both with R$_{ext}\sim$10$^{7}$A/W on Si/SiO$_{2}$, see Table 1. However, our R$_{ext}$ is the highest amongst fibre-based PDs for wearable applications, table \ref{table1}.

The temporal photocurrent response is measured in Fig.\ref{fig8}f via an digital oscilloscope (MSO9404A). This gives a rise time$\sim$5ms and a fall time$\sim$35ms, faster than the$\sim$200ms reported for carbon fibre/PVK PDs\cite{sun2018ultrahigh} and$\sim$40ms for fibre based P3HT/ZnO PDs\cite{du2022piezo}, and comparable to PDs made of mechanically exfoliated SLG and QDs on Si/SiO$_{2}$\cite{konstantatos2012hybrid}.

The specific detectivity (D$^{*}$) ($(cm.Hz^{1/2}/W)$ or Jones) relates the performance of PDs in terms of R$_{ext}$ to the PD photoactive area, allowing the comparison of PDs with different active areas\cite{sze2006physics}:
\begin{equation}
D^* =\frac{(AB)^{1/2}}{NEP}
\end{equation}
where A is the photoactive area (projected incident light on fibre), B is the electrical bandwidth (Hz) and NEP($W/Hz^{1/2}$) is the noise equivalent power (the power that gives a signal-to-noise ratio of one in a 1Hz output bandwidth) defined as\cite{fang2019accurate}:
\begin{equation}
NEP = \frac{i_n}{R_{ext}}
\end{equation}
where ${i_n}$ is the dark noise current. The fibre PD is DC biased with a source meter (Keithley 2612B). A chopper is used in the path of the incident beam to modulate the light\cite{fang2019accurate}. The noise (A/$\sqrt{Hz}$) is characterized in the time domain collecting the trace on a digital oscilloscope, with subsequent Fourier transform to analyze the data in the spectral domain.
\begin{figure}
\centerline{\includegraphics[width=90mm]{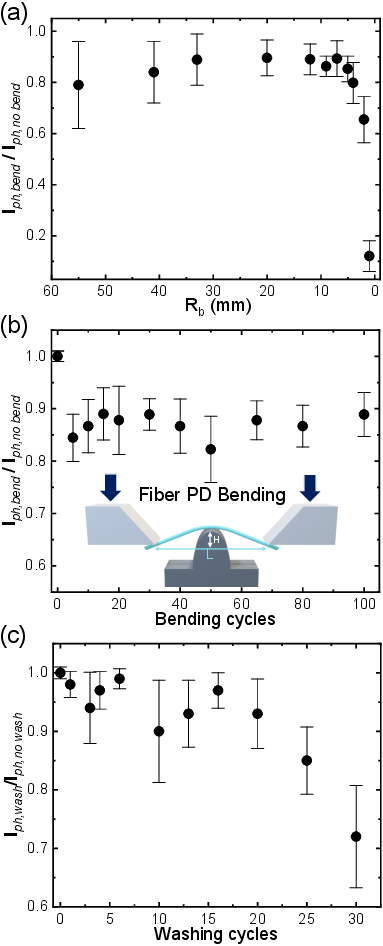}}
\caption{(a)I$_{ph,bend}$/I$_{ph,no bend}$ for N$_{R}$=6 as a function of R$_{b}$. (b) I$_{ph,bend}$/I$_{ph,no bend}$ for N$_{R}$=6 as a function of bending cycles for R$_{b}$=4mm. (c) I$_{ph,wash}$/I$_{ph,no wash}$ for N$_{R}$=6 as a function of AATCC washing cycles.}
\label{fig10}
\end{figure}

Fig.\ref{fig9}a plots the spectral noise density as a function of frequency $f$. The measured noise in the dark shows a 1/$f$ dependence. Thus, the noise density (i.e. noise power per unit of bandwidth (dbm.Hz$^{-1/2}$))\cite{sze2006physics} is inversely proportional to $f$, due to charge traps and defects\cite{sze2006physics}. Considering the measurement at 4Hz,$\sim$10 times less than the cut off $f$, defined as the $f$ at which the detector R$_{ext}$ decreases by 3dB\cite{sze2006physics}, we get NEP$\sim$2.1$\times$10$^{-11}$W/Hz$^{1/2}$, D*$\sim$2$\times$10$^9$Jones for N$_R$=6. The noise current in the shot noise limit (due to generation and recombination of e-h pairs and resistive current paths in PDs[6]), can be expressed as $i_n=(2qI_{dark})^{1/2}$\cite{fang2019accurate, bianconi2021exaggerated}, where the combined dark current ($I_{dark}$) is the sum of unamplified graphene current ($I_{dark(SLG)}$) and amplified injection current ($I_{dark(PVK)}$) from PVK, caused by the thermal excitations of charge carriers in the dark. $I_{dark(PVK)}$, which is due to thermal excitations of charge carriers in the dark, is much smaller than $I_{dark(SLG)}\sim$3.75mA. Therefore, following Ref.\cite{bianconi2021exaggerated}, the shot noise limited noise current can be expressed as $i_n=(2qI_{dark(SLG)})^{1/2}$. D* in the shot noise limit can be written as\cite{fang2019accurate}:
\begin{equation}
D^* =\frac{ R_{ext} (AB)^{1/2}}{i_n}
\end{equation}
We get D$^{*}\sim$10$^{13}$ Jones for N$_R$=6. Our D$^{*}$ is comparable with that of state-of-the-art fiber based PDs\cite{sun2018ultrahigh}, where D$^{*}\sim2.15\times10^{13}$ Jones, but with R$_{ext}\sim$562.9mA/W at 800nm and $\tau_{res}\sim$200ms\cite{sun2018ultrahigh}, see Table 1.

Fig.\ref{fig9}b plots R$_{ext}$ as a function of wavelength from 490 up 870nm. This follows the absorption of SLG/PVK films, Fig.\ref{fig6}a, suggesting PVK is the main light absorbing layer in which photoexcited charges are supplied\cite{dou2014solution}. To get a better understanding of the spectral response of PDs versus wavelength, optical simulations are performed by extracting the PVK index from Ref.\cite{tejada2021hybrid}. We use the PVK refractive index in transfer matrix calculations to estimate the absorption of SLG/PVK on our fibre structure. The imaginary and the real part of the PVK average index is found by applying the Kramers-Kronig (KK) relation, $n_{PVK}(w)$ = 1 + 2$\pi^{-1}$ $\mathcal{P}$ $\int_{0}^{\infty}\, w^{'}\, K_{PVK}(w^{'})/(w^{'2}-w^{2})\, dw^{'}$, where $\mathcal{P}$ denotes the principal value of the integral and $w$ is the angular $f$. The SLG refractive index is modelled by the Kubo conductance\cite{hanson2008quasi} at RT and E$_{F}$=380meV. The spectral response of the fibre PDs in Fig.\ref{fig9}b follows the optical absorption of SLG/PVK on fibre in Fig.\ref{fig9}b, i.e. drop of R$_{ext}$ and absorption with increasing wavelength.

Mechanical characterizations are crucial to emulate real-life applications of wearable PDs\cite{cai2019materials}. These include washing in a tumble washing machine\cite{zeng2014fiber, mattmann2008sensor} and deformations against body motions (such as bending\cite{zeng2014fiber}). We thus test our fibre-based PDs according to AATCC\cite{AATCC} for washing. For mechanical deformations, we perform bending cycles, starting from a radius of curvature of 55mm where the PDs show almost no performance change, to 1mm where we observe PD degradation.

Fig.\ref{fig10}b is the schematic of a three-point bending setup (Deben Microtest) we use to characterize the effect of bending on the PD photocurrent. The photocurrent is first measured for no bending (I$_{ph,no bend}$), and then under bending (I$_{ph,bend}$). We use the ratio of (I$_{ph,bend}$) over (I$_{ph,no bend}$) to monitor the effect of mechanical deformations. The bending radius $(R_b)$ is calculated as\cite{hearle1969structural}:
\begin{equation}
R_b = \frac{[H^2+(L/2)^2]}{2H}
\end{equation}
where $H$ is the height at the chord midpoint of two ends of the arc, and $L$ is the chord of circumference connecting two ends of the arc.

Fig. 10 plots average and standard deviations (error bars) of I$_{ph,bend}$/I$_{ph,no bend}$ as a function of R$_b$ and bending cycles, as well as I$_{ph,wash}$/I$_{ph,no wash}$ as a function of washing cycles. I$_{ph,bend}$/I$_{ph,no bend}$ shows up to$\sim$20\% variation for bending radius as low as 4mm (R$_b$=4mm), Fig.\ref{fig10}a. Fig.10 suggests that variations primarily occur during the initial cycles. Refs.\cite{Qiaoadd, Panadd, Yangadd} reported that the electrical behaviour of SLG on flexible substrates (e.g. PET, PDMS) stabilizes after several deformations. Ref.\cite{Qiadd} reported that the morphology of SLG wrinkles undergoes gradual modification with increasing bending, leading to a change of contact area. We attribute the slight increase of  I$_{ph,bend}$/I$_{ph,no bend}$ up to R$_b\sim$4mm to changes in contact area of rolled CVD SLG on PVK. Upon further bending (R$_b<$4mm), I$_{ph,bend}$ drops to$\sim$10\% of I$_{ph,no bend}$, due to failure of various components, such as contact of Ag electrodes with graphene and crack formation in the fibre PD. Despite pushing our PDs to much lower R$_{b}$ (4mm) compared to InP-based semiconductor flexible PDs (R$_{b}>$38.1mm)\cite{yang2010large}, we achieve five orders of magnitude higher R$_{ext}$ (22,000A/W here vs 0.12A/W at 533nm\cite{yang2010large}), Table 1. Ref.\cite{dang2016methylammonium} reported hybrid CVD SLG and PVK on Polyimide with R$_{b}$=12mm\cite{dang2016methylammonium}, one order of magnitude lower R$_{ext}$=115A/W at 515 nm\cite{dang2016methylammonium} compared to ours. Ref.\cite{chen2016flexible} fabricated PVK/Conjugated-Polymer composite PDs on PET with comparable R$_b$ (R$_b=$4mm), yet, the latter has six orders of magnitude lower R$_{ext}\sim$154mA/W at 835nm\cite{chen2016flexible}, see Table 1.

We then test our devices for a fixed R$_b$=4mm (as we observe$\sim$20\% variation in I$_{ph,bend}$/I$_{ph,no bend}$ upto R$_b$=4mm in Fig.\ref{fig10}a) as a function of bending cycles, Fig.\ref{fig10}b. After 100 bending cycles (I$_{ph,bend}$)/I$_{ph,no bend}$ changes$<$20\%, Fig.\ref{fig10}b. In similar conditions, Ref.\cite{sun2018ultrahigh} reported 6 fold lower R$_{ext}\sim$560mA/W at 800nm\cite{sun2018ultrahigh} for 60 less cycles, see Table 1.

We then test the PDs in a tumble washing machine (SKYLINE rotate washing color) at 40$\degree$C up to 30 washing cycles, each programmed for 45min, total volume$\sim$0.37\% (Persil detergent powder), and with ten 6mm stainless steel balls\cite{AATCC}. Two encapsulation approaches are used to protect the PDs during washing. The fibres are conformally coated with$\sim$1$\mu$m parylene C, since this acts as a barrier to moisture\cite{fukuda2010thermal} and gases\cite{fukuda2010thermal}. The PDs are then further coated with PDMS for additional waterproofing and resistance against tension/compression introduced during washing. Part A and B of the commercial Si-based elastomer Polycraft T15 are mixed 1:1 using a speed mixer (DAC 150 FVZ-K) at 2000rpm for 30s. The resulting solution is then spread on the fibre and cured at 80$^{\circ}$C for 30min to expedite the process and ensure crosslinking of polymer chains\cite{park2012three}. After encapsulation, the devices are tested at 488nm at 1V. They have R$_{ext}\sim$22,000A/W, with no degradation after encapsulation. The photocurrent of fibre-based PDs is first measured for no washing step (I$_{ph,no wash}$). The PDs are then placed in a tumble washing machine and washed. The PDs are then drip dried in a ventilated oven at 60$^{\circ}$C (Genlab Air), then measured at various washing cycles (I$_{ph,wash}$) from 1 to 30 cycles, Fig.\ref{fig10}c. After 20 cycles, the PDs show$\sim$6\% degradation in I$_{ph,wash}$/I$_{ph,no wash}$, Fig.\ref{fig10}c. After 30 washing cycles, I$_{ph,wash}$/I$_{ph,no wash}$ drops$\sim$28\%, due to degradation of electrodes caused by stresses during washing. Ref.\cite{carey2017fully} reported$\sim$50\% drop in drain current after 10 and 20 washing cycles, however, our fibre PDs maintained up to$\sim$72\% of the I$_{ph}$ after 30 washing cycles.
\section{\label{sec:level3}Conclusions}
We reported fibre-based SLG/PVK PDs with $R_{ext}$ up to$\sim$22,000A/W at 1V operating voltage at 488nm. The PDs have a broad spectral responsivity between 488 and 870nm, with rise and fall times$\sim$5ms and$\sim$35ms. The devices were subjected to mechanical bending tests (100 cycles at R$_{b}\sim$4mm) with I$_{ph,bend}$/I$_{ph,no bend}$ changes$<$10\% and washing tests (AATCC standard) with I$_{ph,wash}$/I$_{ph,no wash}\sim$94\% and$\sim$72\% at 20 and 30 cycles, respectively. The combination of high ($\sim$22KA/W) responsivity, response time, flexibility, washability, and low ($<$1V) operating voltage make our PDs attractive for wearable applications.
\section{\label{sec:level4}Acknowledgments}
We acknowledge funding from EU Graphene Flagship, ERC Grants Hetero2D, GIPT, EU Grants CHARM, Graph-X, GreenCAP, EPSRC Grants EP/K01711X/1, EP/K017144/1, EP/N010345/1, and EP/L016087/1, EP/X015742/1, EP/V000055/1, the Royal Society and Tata Group (UF150033).For open access, we applied a Creative Commons Attribution (CC BY) license to any Author Accepted Manuscript version arising from this submission.

\end{document}